\documentclass[prd,aps,showpacs,nofootinbib]{revtex4}%
\usepackage{amsmath}
\usepackage{amsfonts}
\usepackage{amssymb}
\usepackage{bm}
\usepackage{graphicx}%
\providecommand{\U}[1]{\protect\rule{.1in}{.1in}}
\begin{document}
\title{Dimensional reduction of the CPT-even electromagnetic sector of the Standard
Model Extension}
\author{Rodolfo Casana, Eduardo S. Carvalho, Manoel M. Ferreira Jr}
\affiliation{Departamento de F\'{\i}sica, Universidade Federal do Maranh\~{a}o (UFMA),
Campus Universit\'{a}rio do Bacanga, S\~{a}o Lu\'{\i}s - MA, 65085-580, Brazil.}

\begin{abstract}
The CPT-even abelian gauge sector of the Standard Model Extension is
represented by the Maxwell term supplemented by $\left(  K_{F}\right)
_{\mu\nu\rho\sigma}F^{\mu\nu}F^{\rho\sigma}$, where the Lorentz-violating
background tensor, $\left(  K_{F}\right)  _{\mu\nu\rho\sigma}$, possesses the
symmetries of the Riemann tensor. In the present work, we examine the planar
version of this theory, obtained by means of a typical dimensional reduction
procedure to $(1+2)$ dimensions. The resulting planar electrodynamics is
composed of a gauge sector containing six Lorentz-violating coefficients, a
scalar field endowed with a noncanonical kinetic term, and a coupling term
that links the scalar and gauge sectors. The dispersion relation is exactly
determined, revealing that the six parameters related to the pure
electromagnetic sector do not yield birefringence at any order. In this model,
the birefringence may appear only as a second order effect associated with the
coupling tensor linking the gauge and scalar sectors.The equations of motion
are written and solved in the stationary regime. The Lorentz-violating
parameters do not alter the asymptotic behavior of the fields but induce an
angular dependence not observed in the Maxwell planar theory.

\end{abstract}

\maketitle

\section{Introduction}

Lorentz symmetry has been considered as a fundamental cornerstone of physics
since the establishment of the special theory of relativity. The inquiry about
to what extent this is an exact symmetry of nature constitutes the scope of
most investigations in Lorentz violation nowadays. A motivation for such
studies is that the observation of Lorentz symmetry small violations in
current low-energy phenomena could indicate new routes for developing theories
at the Planck-scale. The Standard Model Extension (SME) \cite{Colladay},
\cite{Samuel} is a theoretical framework that incorporates Lorentz-violating
coefficients to the standard model and to general relativity, and has served
as a suitable tool for constructing interesting approaches in this area.

The gauge sector of the SME embraces twenty three\ Lorentz-violating
coefficients that yield some unconventional phenomena such as vacuum
birefringence and Cherenkov radiation. The coefficients are usually classified
in accordance with some criteria. One criterion is the possible violation of
the CPT symmetry, being the parameters CPT-odd or CPT-even. The
Carroll-Field-Jackiw (CFJ) term \cite{Jackiw} is the CPT-odd term of the SME,
composed of four parameters, that engenders a parity-odd and birefringent
electrodynamics whose properties were largely examined in connection with many
diverse issues: consistency aspects and modifications induced in QED
\cite{Adam,Soldati,Higgs}, supersymmetry \cite{Susy}, generation by radiative
correction \cite{Radio}, vacuum Cherenkov radiation emission \cite{Cerenkov1},
electromagnetic propagation in waveguides \cite{winder2}, Casimir effect
\cite{Casimir}, finite-temperature contributions and Planck distribution
\cite{winder1, Tfinite}, anisotropies of the Cosmic Microwave Background
Radiation\cite{CMBR}, classical electrodynamics solutions\cite{Classical},
dimensional reduction \cite{Dreduction}, \cite{Manojr2}, \cite{Dreduction2}.

The CPT-even gauge sector of the SME has been studied since 2002, after the
pioneering contributions by Kostelecky \& Mewes \cite{KM1,KM2}. This sector is
represented by the nineteen components of the\ fourth-rank tensor, $\left(
K_{F}\right)  _{\alpha\nu\rho\varphi}$, endowed with the same symmetries of
the Riemann tensor, and a double null trace. The nineteen components are
grouped in two subclasses: the ten birefringent ones, which are severely
constrained by astrophysical tests of birefringence, and the nine
nonbirefringent ones. This latter group can only be constrained by laboratory
tests, which are continuously being proposed and realized. High-quality
cosmological spectropolarimetry data \cite{KM3} have been employed to impose
stringent upper bounds $\left(  \text{as tight as }10^{-37}\right)  $ on the
ten birefringent LIV parameters. On the other hand, the Cherenkov radiation
\cite{Cherenkov2} and the absence of emission of Cherenkov radiation by UHECR
(ultrahigh energy cosmic rays) \cite{Klink2,Klink3} have been used to impose
upper bounds on the nonbirefringent components. Photon-fermion vertex
corrections induced by the LIV coefficients \cite{Interac1,Interac2,Interac3},
\cite{Bocquet} have been employed to state upper bounds on these coefficients,
as well.

The dimensional reduction of the CPT-odd gauge term of Standard Model
Extension was performed in Ref. \cite{Dreduction}, yielding a planar
electrodynamics composed of the Maxwell-Chern-Simons electrodynamics coupled
with a massless scalar field - the remanent of the third spacial component of
the four-potential $\left(  A^{\left(  3\right)  }=\phi\right)  $. It is
interesting to note that the Chern-Simons terms appears naturally in such
reduction. This planar model was studied in its consistency (stability,
causality and unitarity) and had its classical solutions determined in Ref.
\cite{Manojr2}. The dimensional reduction of the Abelian-Higgs
Maxwell-Carroll-Field-Jackiw model was performed in Ref. \cite{Dreduction2}.

In the present work, we realize the dimensional reduction of the CPT-even
gauge sector of the SME to $(1+2)-$dimensions following the prescription
adopted in Refs.\cite{Dreduction},\cite{Dreduction2}, that is, freezing the
third spatial component in such a way the fields can not exhibit any
dependence on it. The arising scalar field is the remanent of the third
spacial component of the four-potential $\left(  A^{\left(  3\right)  }%
=\phi\right)  $. We obtain a planar theory composed of the electromagnetic
sector, a scalar massless field with noncanonical kinetic term, and a mixing
term that couples the scalar and gauge sectors. In the gauge sector, Lorentz
violation is induced by a fourth-rank tensor, $Z_{\mu\nu\lambda\kappa},$
endowed with the symmetries of the Riemann tensor, which renders six
independent components. The scalar sector presents an additional noncanonical
kinetic term, $C_{\mu\lambda}{\partial^{\mu}\phi\partial^{\lambda}\phi,}$
where $C_{\mu\lambda}$ is a Lorentz-violating symmetric second-rank tensor
with six independent components. The scalar and gauge sectors are coupled by
the\ third-rank tensor, $T_{\mu\lambda\kappa},$ whose symmetries imply eight
independent components. The traceless condition, coming from the (1+3)
dimensional model, reduced the number of independent parameters to nineteen.
Once the planar model and its structural features are set up, we examine the
effects of the Lorentz-violating parameters in the electromagnetic classical
solutions (in the stationary regime), using the Green's method. As in the
original four-dimensional counterpart, stationary currents and static charge
are able to create both magnetic and electric fields in this planar theory.
The stationary solutions reveal that Lorentz-violating coefficients do not
modify the asymptotic radial behavior of the Maxwell planar electrodynamics,
but are able to generate terms with angular dependence. The dispersion
relation stemming from the pure gauge sector allows to notice that the planar
electrodynamics is free from birefringence, which is implied only at second
order by the components of the tensor $T_{\mu\lambda\kappa}$.

This work is organized as follows. In Sec.II, we briefly present general
features of the CPT-even electrodynamics of the SME. In Sec.III, we perform
the dimensional reduction procedure that leads to the planar theory of
interest. In Sec. IV, we study this planar theory focusing on the equations of
motion and the attainment of the dispersion relation. In Sec. V, we evaluate
the classical stationary solutions for electric and magnetic field, remarking
the deviations induced by the Lorentz-violating terms. In Sec. VI, we present
our\ conclusions and final remarks.

\section{The parametrization of the CPT-even gauge sector of SME in (1+3)
dimensions}

The CPT-even sector of the Lorentz-violating electrodynamics of the SME photon
sector is represented by the following Lagrangian:
\begin{equation}
\mathit{{\mathcal{L}}}=-\frac{1}{4}F_{\hat{\mu}\hat{\nu}}F^{\hat{\mu}\hat{\nu
}}-\frac{1}{4}\left(  K_{F}\right)  _{\hat{\mu}\hat{\nu}\hat{\lambda}%
\hat{\kappa}}F^{\hat{\mu}\hat{\nu}}F^{\lambda\kappa}-J_{\hat{\mu}}A^{\hat{\mu
}}, \label{L1}%
\end{equation}
where the indices with hat, $\hat{\mu},\hat{\nu},$ run from 0 to 3,
$F_{\hat{\mu}\hat{\nu}}$ is the usual electromagnetic field tensor,
$A^{\hat{\mu}}$ is the four-potential, $\left(  K_{F}\right)  _{\hat{\mu}%
\hat{\nu}\hat{\lambda}\hat{\kappa}}$ is a renormalizable, dimensionless
coupling which has the same symmetries as the Riemann tensor
\begin{equation}
\left(  K_{F}\right)  _{\hat{\mu}\hat{\nu}\hat{\lambda}\hat{\kappa}}=-\left(
K_{F}\right)  _{\hat{\nu}\hat{\mu}\hat{\lambda}\hat{\kappa}},~\ \ \ \left(
K_{F}\right)  _{\hat{\mu}\hat{\nu}\hat{\lambda}\hat{\kappa}}=-\left(
K_{F}\right)  _{\hat{\mu}\hat{\nu}\hat{\kappa}\hat{\lambda}},~\ \ \ \left(
K_{F}\right)  _{\hat{\mu}\hat{\nu}\hat{\lambda}\hat{\kappa}}=\left(
K_{F}\right)  _{\hat{\lambda}\hat{\kappa}\hat{\mu}\hat{\nu}},
\end{equation}%
\begin{equation}
\left(  K_{F}\right)  _{\hat{\mu}\hat{\nu}\hat{\lambda}\hat{\kappa}}+\left(
K_{F}\right)  _{\hat{\mu}\hat{\lambda}\hat{\kappa}\hat{\nu}}+\left(
K_{F}\right)  _{\hat{\mu}\hat{\kappa}\hat{\nu}\hat{\lambda}}=0.
\end{equation}
and a double null trace, $\left(  K_{F}\right)  ^{\hat{\mu}\hat{\nu}}{}%
_{\hat{\mu}\hat{\nu}}=0.$ The equation of motion is%
\begin{equation}
\partial_{\hat{\nu}}F^{\hat{\nu}\hat{\mu}}-\left(  K_{F}\right)  ^{\hat{\mu
}\hat{\nu}\hat{\lambda}\hat{\kappa}}{}\partial_{\hat{\nu}}F_{\hat{\lambda}%
\hat{\kappa}}=0. \label{Mot1}%
\end{equation}
The tensor $\left(  K_{F}\right)  _{\alpha\nu\rho\varphi}$ has 19 independent
components, from which nine do not yield birefringence. A very useful
parametrization for addressing this theory is the one presented in Refs.
\cite{KM1,KM2}, in which these 19 components are contained in four $3\times3$
matrices:
\begin{align}
\left(  \widetilde{\kappa}_{e+}\right)  ^{jk}  &  =\frac{1}{2}(\kappa
_{DE}+\kappa_{HB})^{jk},~~\left(  \widetilde{\kappa}_{e-}\right)  ^{jk}%
=\frac{1}{2}(\kappa_{DE}-\kappa_{HB})^{jk}-\frac{1}{3}\delta^{jk}(\kappa
_{DE})^{ii},~~\kappa_{\text{tr}}=\frac{1}{3}\text{tr}(\kappa_{DE}),~~\\
\left(  \widetilde{\kappa}_{o+}\right)  ^{jk}  &  =\frac{1}{2}(\kappa
_{DB}+\kappa_{HE})^{jk},\text{ ~~}\left(  \widetilde{\kappa}_{o-}\right)
^{jk}=\frac{1}{2}(\kappa_{DB}-\kappa_{HE})^{jk},
\end{align}
where $\widetilde{\kappa}_{e}$ and $\widetilde{\kappa}_{o}$ designate
parity-even and parity-odd matrices, respectively. The $3\times3$ matrices
$\kappa_{DE},\kappa_{HB},\kappa_{DB},\kappa_{HE}$ are defined in terms of the
$\left(  K_{F}\right)  -$tensor components:
\begin{align}
\left(  \kappa_{DE}\right)  ^{jk}  &  =-2\left(  K_{F}\right)  ^{0j0k},\text{
}\left(  \kappa_{HB}\right)  ^{jk}=\frac{1}{2}\epsilon^{jpq}\epsilon
^{klm}\left(  K_{F}\right)  ^{pqlm},\label{P1}\\
\text{ }\left(  \kappa_{DB}\right)  ^{jk}  &  =-\left(  \kappa_{HE}\right)
^{kj}=\epsilon^{kpq}\left(  K_{F}\right)  ^{0jpq}. \label{P2}%
\end{align}
The matrices $\kappa_{DE},\kappa_{HB}$ contain together 11 independent
components while $\kappa_{DB},\kappa_{HE}$ possess together 8 components,
which sums the 19 independent elements of the tensor $\left(  K_{F}\right)
_{\alpha\nu\rho\varphi}$. From these 19 coefficients, 10 are sensitive to
birefringence and 9 are nonbirefringent. These latter ones are contained in
the matrices $\widetilde{\kappa}_{o+}$ and $\widetilde{\kappa}_{e-}.$ The
analysis of birefringence data reveals the coefficients of the matrices
$\widetilde{\kappa}_{e+}$ and $\widetilde{\kappa}_{o-}$ are bounded to the
level of 1 part in $10^{32}$ \cite{KM1,KM2} and 1 part in $10^{37}$\cite{KM3}.

\section{The Dimensional reduction of the CPT-even sector}

In order to study this model in $(1+2)-$dimensions, one realizes its
dimensional reduction, which consists effectively in adopting the following
ansatz over any 4-vector: (i) one keeps unaffected the temporal and also the
first two spatial components; (ii) one freezes the third spacial dimension by
splitting it from the body of the new 3-vector and requiring that the new
quantities $\left(  \chi\right)  $, defined in $(1+2)-$dimensions, do not
depend on the third spacial dimension: $\partial_{_{3}}\chi\longrightarrow0.$
This procedure was performed for the Carroll-Field-Jackiw electrodynamics in
Ref. \cite{Dreduction}. Applying this prescription to the gauge 4-vector,
$A^{\mu}$, one has:
\begin{equation}
A^{\hat{\nu}}\longrightarrow(A^{\nu};\phi),
\end{equation}
where $A^{\left(  3\right)  }=\phi$ is now a scalar field and the Greek
indices without hat run from $0$ to $2$, that is $\mu=0,1,2$. Carrying out
this prescription for the terms of \ Lagrangian (\ref{L1}), one then obtains:
\begin{equation}
F_{\hat{\mu}\hat{\nu}}F^{\hat{\mu}\hat{\nu}}=F_{\mu\nu}F^{\mu\nu}+2F_{\mu
3}F^{\mu3}=F_{\mu\nu}F^{\mu\nu}-2\partial_{\mu}\phi\partial^{\mu}\phi,
\label{F1}%
\end{equation}
\begin{equation}
\left(  K_{F}\right)  _{\hat{\mu}\hat{\nu}\hat{\lambda}\hat{\kappa}}%
F^{\hat{\mu}\hat{\nu}}F^{\hat{\lambda}\hat{\kappa}}=Z_{\mu\nu\lambda\kappa
}F^{\mu\nu}F^{\lambda\kappa}+2Z_{\mu3\lambda\kappa}F^{\mu3}F^{\lambda\kappa
}+2Z_{\mu\nu\lambda3}F^{\mu\nu}F^{\lambda3}+4Z_{\mu3\lambda3}F^{\mu
3}F^{\lambda3},
\end{equation}
where $Z_{\mu\nu\lambda\kappa}$ is the planar version of the original $\left(
K_{F}\right)  $-tensor, that is, $Z_{\mu\nu\lambda\kappa}=\left[  \left(
K_{F}\right)  _{\mu\nu\lambda\kappa}\right]  _{1+2}.$ It fulfills the
following symmetry properties%
\begin{equation}
Z_{\mu\nu\lambda\kappa}=Z_{\lambda\kappa\mu\nu},\text{ }Z_{\mu\nu\lambda
\kappa}=-Z_{\nu\mu\lambda\kappa},\text{ \ }Z_{\mu\nu\lambda\kappa}=-Z_{\mu
\nu\kappa\lambda},
\end{equation}
\begin{equation}
Z_{\mu\nu\lambda\kappa}+Z_{\mu\lambda\kappa\nu}+Z_{\mu\kappa\nu\lambda}=0.
\label{Perm}%
\end{equation}

These symmetries imply $Z_{\mu\nu\lambda3}F^{\mu\nu}F^{\lambda3}%
=Z_{\mu3\lambda\kappa}F^{\mu3}F^{\lambda\kappa},$ leading to%
\begin{align}
\left(  K_{F}\right)  _{\hat{\mu}\hat{\nu}\hat{\lambda}\hat{\kappa}}%
F^{\hat{\mu}\hat{\nu}}F^{\hat{\lambda}\hat{\kappa}} &  =Z_{\mu\nu\lambda
\kappa}F^{\mu\nu}F^{\lambda\kappa}+4Z_{\mu3\lambda\kappa}F^{\mu3}%
F^{\lambda\kappa}+4Z_{\mu3\lambda3}F^{\mu3}F^{\lambda3},\\
\left(  K_{F}\right)  _{\hat{\mu}\hat{\nu}\hat{\lambda}\hat{\kappa}}%
F^{\hat{\mu}\hat{\nu}}F^{\hat{\lambda}\hat{\kappa}} &  =Z_{\mu\nu\lambda
\kappa}F^{\mu\nu}F^{\lambda\kappa}-4T_{\mu\lambda\kappa}\partial^{\mu}\phi
F^{\lambda\kappa}+4C_{\mu\lambda}\partial^{\mu}\phi\partial^{\lambda}\phi,
\end{align}
where $F^{\mu3}=\partial^{\mu}\phi,$ and it were defined new second-rank and
third-rank tensors
\[
T_{\mu\lambda\kappa}=\left(  K_{F}\right)  _{3\mu\lambda\kappa},\text{ }%
C_{\mu\lambda}=\left(  K_{F}\right)  _{\mu3\lambda3}.
\]
With it, the dimensionally reduced Lagrangian is
\begin{equation}
\mathit{{\mathcal{L}}}_{(1+2)}=-\frac{1}{4}F_{\mu\nu}F^{\mu\nu}-\frac{1}%
{4}Z_{\mu\nu\lambda\kappa}F^{\mu\nu}F^{\lambda\kappa}+\frac{1}{2}\partial
_{\mu}\phi\partial^{\mu}\phi-C_{\mu\lambda}{\partial^{\mu}\phi\partial
^{\lambda}\phi}+T_{\mu\lambda\kappa}{\partial^{\mu}\phi}F^{\lambda\kappa
}-J_{\mu}A^{\mu}-J\phi,\label{LP1}%
\end{equation}
The presence of the tensor $C_{\mu\lambda}$ provides a noncanonical kinetic
term for the scalar field. Some attempts of proposing Lorentz-violating
constructions for topological defects with a term like this are already known
in literature \cite{LV}. This term has recently been used to study acoustic
black holes with Lorentz-violation in (1+2) dimensions \cite{Brito} and also
the Bose-Einstein condensation of a bosinic ideal gas \cite{Casana}. The
tensor $T_{\mu\lambda\kappa},$ in turn, is responsible for the coupling
between the scalar and gauge sectors in this planar theory. These two tensors
satisfy the following symmetries:%
\begin{align}
C_{\mu\lambda} &  =C_{\lambda\mu},\label{sym1}\\
\text{ }T_{\mu\lambda\kappa} &  =-T_{\mu\kappa\lambda}.\label{sym2}\\
T_{\mu\lambda\kappa}+T_{\lambda\kappa\mu}+T_{\kappa\mu\lambda} &
=0,\label{sym3}%
\end{align}
The (double) traceless property of the $K_{F}-$tensor is now read as
\begin{equation}
Z_{\mu\nu}^{\text{ \ \ \ }\mu\nu}+2C_{\text{ \ }\alpha}^{\alpha}%
=0.\label{trace}%
\end{equation}
By the relations (\ref{sym1}-\ref{sym3}), we conclude that the tensors
$C_{\mu\lambda},$ $Z_{\mu\nu\lambda\kappa},$ $T_{\mu\lambda\kappa}$ contain
six, six and eight independent components, respectively, comprising twenty
parameters. The relation (\ref{trace}) states a constraint between them,
remaining nineteen independent components, the same number of the tensor
$K_{F}$ (before the dimensional reduction).

The reduced model (\ref{LP1}) is invariant under the following local gauge
transformation:%
\begin{equation}
A_{\mu}\rightarrow A_{\mu}+\partial_{\mu}\Lambda,~\ \ \phi\rightarrow\phi,
\end{equation}
in such a way it preserves the $U\left(  1\right)  $ local gauge symmetry of
the 4-dimensional model. The full Lagrangian of this model is written as
\begin{align}
\mathit{{\mathcal{L}}}  &  =-\frac{1}{4}F_{\mu\nu}F^{\mu\nu}+\frac{1}%
{2}\partial_{\mu}\phi\partial^{\mu}\phi-C_{00}\left(  \partial_{0}\phi\right)
^{2}{+}C_{0i}\left(  \partial_{0}\phi\right)  (\partial_{i}\phi)-C_{ij}\left(
\partial_{i}\phi\right)  (\partial_{j}\phi)\nonumber\\
&  -(Z_{0i12}E^{i}\mathbf{)}B-\frac{1}{2}\left(  k_{DE}\right)  _{ij}%
E^{i}E^{j}-\frac{1}{2}sB^{2}-T_{00i}\partial_{0}\phi E^{i}-\epsilon
_{ij}T_{0ij}\partial_{0}\phi B+T_{i0j}\partial_{i}\phi E^{j}+\epsilon
_{lj}T_{ilj}\partial_{i}\phi B.
\end{align}
As in the original four-dimensional model, the Lorentz-violating coefficients
have definite parity. In (1+2) dimensions, the parity operator acts doing
$r\rightarrow(-x,y)$, so that the fields go as
\begin{equation}
A_{0}\rightarrow A_{0},\mathbf{\ }A\rightarrow(-A_{x},A_{y}),\mathbf{\ }%
E\rightarrow(-E_{x},E_{y}),\mathbf{\ }B\rightarrow-B\mathbf{.\ }%
\end{equation}
For more details, see Ref. \cite{Planar}. We consider that the field $\phi$
behaves as a scalar, $\phi\rightarrow\phi.$ This allows to conclude that this
planar model possesses twelve parity-even components, and nine parity-odd
ones, as shown in the Table I. Further, we see that the trace relation
(\ref{trace}) involves only parity-even coefficients, whereas the relation
(\ref{sym3}) embraces only parity-odd parameters (when the indices of the
tensor $T_{\kappa\mu\lambda}$ assume three different values, $T_{012}%
+T_{120}+T_{201}=0$). These two relations reduce the number of independent
components from twenty one to nineteen, as it is expected. The fact that the
components of the vectors $\left(  Z_{0i12},T_{00i}\right)  $ transform
distinctly is a consequence of the way the vectors $\mathbf{r},$\textbf{
}$\mathbf{A},$\textbf{ }$\mathbf{E}$ behave under parity\footnote{Note that in
the case the field $\phi\ $behaves like a pseudoscalar $\left(  \phi
\rightarrow-\phi\right)  ,$ the behavior of the components $T_{00i}%
,T_{0ij},T_{i0j},T_{ilj}$ is reversed under parity.}.\textbf{ }%
\begin{table}[h]
\ \ \ \ \ \
\begin{tabular}
[c]{|l|l|l|l|}\hline
Components & \multicolumn{1}{|l|}{\ } & N & \ $\mathbb{N} $\\\hline
\ Parity-even & $C_{00},C_{02},C_{11},C_{22},L_{1},\left(  k_{DE}\right)
_{11},\left(  k_{DE}\right)  _{22},s,T_{002},T_{101},T_{202},T_{112}$ & $12$ &
$11$\\\hline
Parity-odd & $C_{01},C_{12},L_{2},\left(  k_{DE}\right)  _{12},T_{001}%
,T_{012},T_{102},T_{201},T_{212}$ & $9$ & $8$\\\hline
Total & \multicolumn{1}{|l|}{$\ \ \ \ \ \ \ $} & $21$ & $19$\\\hline
\end{tabular}
\caption{Parity-classification and number of Lorentz-violating parameters
belonging to the planar model. The symbol N designates the number of
components, while $\mathbb{N}$ designates the total of independent
components.}%
\label{A_1B}%
\end{table}

If one neglects the coupling between the scalar and gauge sectors $\left(
T_{\mu\lambda\kappa}=0\right)  ,$ one has a planar theory composed by the
usual Maxwell electrodynamics modified by the term $Z_{\mu\nu\lambda\kappa
}F^{\mu\nu}F^{\lambda\kappa}$ and a scalar field endowed with a noncanonical
kinetic term, whose properties will be examined. The planar Lagrangian density
of the electromagnetic sector is
\begin{equation}
\mathit{{\mathcal{L}}}_{EM(1+2)}=-\frac{1}{4}F_{\mu\nu}F^{\mu\nu}-\frac{1}%
{4}Z_{\mu\nu\lambda\kappa}F^{\mu\nu}F^{\lambda\kappa}-J_{\mu}A^{\mu},
\label{LEM}%
\end{equation}
which represents a gauge-invariant theory (in the absence of external
currents). We should note that the planar tensor $Z_{\mu\nu\lambda\kappa}$
would possess $3^{4}=81$ components in the absence of the symmetries. The
symmetries properties, however, reduce them to only six independents
components:
\begin{equation}
Z_{0ilm}=[Z_{0112},Z_{0212}],\text{ }Z_{0i0m}=\left[  Z_{0101},Z_{0202}%
,Z_{0102}\right]  ,\text{ }Z_{ijlm}=\left[  Z_{1212}\right]  .
\end{equation}
It is interesting to note that the permutation symmetry (\ref{Perm}) is now
just a complementary relation, not implying a new constraint on the components
of the tensor $Z_{\mu\nu\lambda\kappa}.$ This planar tensor does not share the
double traceless condition of the tensor $\left(  K_{F}\right)  $ anymore.
Instead of it holds Eq.(\ref{trace}), that states a relation between its
components and the ones of the tensor $C_{\mu\lambda}.$ For this reason, when
the gauge and scalar sector are considered together, both ones only contribute
with eleven components.

In order to propose an effective parametrization for gauge sector elements, it
is helpful to write the Lagrangian element, $Z_{\mu\nu\lambda\kappa}F^{\mu\nu
}F^{\lambda\kappa}$, in terms of the electric and magnetic fields,
\begin{equation}
Z_{\mu\nu\lambda\kappa}F^{\mu\nu}F^{\lambda\kappa}=4Z_{0i12}E^{i}%
B+4Z_{0i0j}E^{i}E^{j}+4Z_{1212}B^{2}, \label{Z1}%
\end{equation}
where it was used $F^{0i}=-E^{i},$ $F^{12}=F_{12}=-B.$ We should remember that
in $(1+2)-$dimensions the magnetic field is a scalar. Thus, the two elements
$Z_{0ilm}$ can be read as elements of a two-vector,%
\begin{equation}
2Z_{0ilm}=2Z_{0i12}=L_{i},\text{ }L^{i}=2Z^{0i12},
\end{equation}
with \ $\mathbf{L}=(L_{1},L_{2}).$ The three elements $Z_{0i0j}$ are written
as elements of a symmetric $2\times2$ matrix%
\begin{equation}
2Z_{0i0j}=\left(  k_{DE}\right)  _{ij}=\left(  k_{DE}\right)  _{ji},
\end{equation}
whose components are
\begin{equation}
k_{DE}=2\left[
\begin{array}
[c]{cc}%
Z_{0101} & Z_{0102}\\
Z_{0102} & Z_{0202}%
\end{array}
\right]  .
\end{equation}
Finally, the single element $Z_{ijlm}$ plays the role of a scalar,
\begin{equation}
2Z_{ijlm}=2Z_{1212}=s.
\end{equation}
Using the new definitions, the planar pure electromagnetic Lagrangian
(\ref{LEM}) takes the form
\begin{equation}
\mathit{{\mathcal{L}}}_{EM(1+2)}=\frac{1}{2}\mathbf{E}^{2}-\frac{1}{2}\left(
k_{DE}\right)  _{ij}E^{i}E^{j}-\frac{1}{2}\left(  1+s\right)  B^{2}%
+(\mathbf{L}\cdot\mathbf{E)}B, \label{Lag2}%
\end{equation}
where it was used the contraction\textbf{ }%
\begin{equation}
Z_{\mu\nu\lambda\kappa}F^{\mu\nu}F^{\lambda\kappa}=-4(\mathbf{L}%
\cdot\mathbf{E)}B+2E^{i}\left(  k_{DE}\right)  _{ij}E^{j}+2sB^{2}.\text{ }%
\end{equation}

Another relevant aspect concerns the evaluation of the canonical
energy-momentum tensor for the planar Lagrangian, (\ref{LEM}), carried out
from the usual form $\Theta^{\beta\rho}=\left[  \partial\mathcal{L}%
/\partial\left(  \partial_{\beta}A_{\alpha}\right)  \right]  \partial^{\rho
}A_{\alpha}-g^{\beta\rho}\mathcal{L}$, and leading to the result,
\begin{equation}
\Theta^{\beta\rho}=-(F^{\beta\alpha}+Z^{\beta\alpha\lambda\kappa}%
F_{\lambda\kappa})\partial^{\rho}A_{\alpha}-g^{\beta\rho}\mathcal{L}.
\end{equation}
The energy density,
\begin{equation}
\Theta_{EM}^{00}=\frac{1}{2}(\mathbf{E}^{2}+B^{2})-\frac{1}{2}\left(
k_{DE}\right)  ^{ij}E^{i}E^{j}+\frac{1}{2}sB^{2}, \label{T_00}%
\end{equation}
is obtained by using the Gauss%
%TCIMACRO{\U{b4}}%
%BeginExpansion
\'{}%
%EndExpansion
s law. It is interesting to mention that the same result is achieved via the
construction of the density of Hamiltonian, $H=\pi^{\alpha}\overset{\cdot
}{A}_{\alpha}-L$, where $\pi^{\alpha}=\partial L/\partial\left(  \partial
_{0}A_{\alpha}\right)  $ is the conjugate momentum,
\begin{equation}
\pi^{\alpha}=-F^{0\alpha}-Z^{0\alpha\lambda\kappa}F_{\lambda\kappa}.
\end{equation}

In components, we have $\pi^{0}=0$ and $\pi^{i}=E^{i}-\left(  k_{DE}\right)
^{ij}E^{j}+L^{i}B.$ The pure gauge model has two first class constraints,
$\pi^{0}$ and $\partial_{i}\pi^{i}$, the latter one being the Gauss's law. The
Hamiltoninan analysis implies the same energy density of Eq. (\ref{T_00}).
These outcomes show that the energy density can be regarded as positive
definite, once the Lorentz-violating parameters are sufficiently small.

\section{Wave equations for the planar electrodynamics}

In order to obtain the classical solutions of the planar electrodynamics
represented by Lagrangian (\ref{LP1}), we should write the equations motion.
In a general way, such equations are given by%
\begin{align}
\partial_{\alpha}F^{\alpha\beta}-Z^{\beta\alpha\lambda\kappa}\partial_{\alpha
}F_{\lambda\kappa}-2T^{\mu\alpha\beta}\partial_{\alpha}\partial_{\mu}\phi &
=J^{\beta},\label{G1}\\
\square\phi+T^{\alpha\lambda\kappa}\partial_{\alpha}F_{\lambda\kappa
}-2C^{\alpha\lambda}\partial_{\alpha}{\partial}_{\lambda}{\phi}  &  {=}{-J.}
\label{S1}%
\end{align}
In the absence of the coupling term $\left(  T^{\mu\alpha\beta}=0\right)  $,
the gauge and scalar sectors become decoupled and classically governed by the
following equation:
\begin{align}
\partial_{\alpha}F^{\alpha\beta}-Z^{\beta\alpha\lambda\kappa}\partial_{\alpha
}F_{\lambda\kappa}  &  =J^{\beta},\label{G2}\\
\square\phi-2C^{\alpha\lambda}\partial_{\alpha}{\partial}_{\lambda}{\phi}  &
={-J,} \label{S2}%
\end{align}
The main reason for neglecting the tensor $T_{\mu\lambda\kappa}$ is that it
appears as a second order contribution in the equations defined in terms only
of the gauge field or the scalar field. In order to verify it, we isolate the
scalar field in Eq. (\ref{S1}), in the absence of scalar sources,
\textbf{$J=0,$} writing%
\begin{equation}
{\phi=}{-}\frac{{T^{\alpha\lambda\kappa}\partial_{\alpha}}}{\square
-2C^{\alpha\lambda}\partial_{\alpha}{\partial}_{\lambda}}{F_{\lambda\kappa}.}
\label{phi1}%
\end{equation}
Replacing Eq. (\ref{phi1}) in Eq. (\ref{G1}), there appears:\textbf{ }%
\begin{equation}
\partial_{\alpha}F^{\alpha\beta}-Z^{\beta\alpha\lambda\kappa}\partial_{\alpha
}F_{\lambda\kappa}+\frac{4T^{\mu\alpha\beta}{T^{\theta\lambda\kappa}%
\partial_{\alpha}\partial_{\mu}\partial_{\theta}}}{\square-2C^{\rho\tau
}\partial_{\rho}{\partial}_{\tau}}{F_{\lambda\kappa}}=J^{\beta}.
\end{equation}
Such expression differs from the decoupled equation (\ref{G2}) by a second
order term in the tensor $T^{\mu\alpha\beta},$ justifying the vanishing choice
$\left(  T^{\mu\alpha\beta}=0\right)  $ adopted. A similar procedure shows
that the tensor $T^{\mu\alpha\beta}$ contributes on the decoupled Eq.
(\ref{S2})\ only at second order, as well. Hence, the gauge and scalar sectors
fulfill decoupled equations of motion at first order, confirming the validity
of Eqs. (\ref{S2}) and (\ref{G2}).

In terms of the electric and magnetic fields, Eq.(\ref{G2}) yields
\begin{equation}
\partial_{i}E^{i}-\left(  k_{DE}\right)  _{ij}\partial_{i}E^{j}+L^{i}%
\partial_{i}B=\rho, \label{Elec1}%
\end{equation}%
\begin{equation}
(1+s)\epsilon^{il}\partial_{l}B-\partial_{t}E^{i}+\left(  k_{DE}\right)
^{ij}\partial_{t}E^{j}-\epsilon^{il}\partial_{l}(L^{j}E^{j})-L^{i}\partial
_{t}B=J^{i}, \label{Mag1}%
\end{equation}
which correspond to modified forms for the Gauss's law and Ampere' law.
Besides these equations, there is the Bianchi identity, $\partial_{\mu}%
F^{\mu\ast}=0,$ where $F^{\mu\ast}=\frac{1}{2}\epsilon^{\mu\nu\alpha}%
F_{\nu\alpha}$ is the the dual of the electromagnetic field tensor in
$(1+2)-$dimensions, which is a three-vector, $F^{\mu\ast}=(-B,-\mathbf{E}%
^{\ast})$. The symbol $(^{\ast})$ designates the dual of a 2-vector: $\left(
E^{i}\right)  ^{\ast}=\epsilon_{ij}E^{j},$ so that $\mathbf{E}^{\ast}%
=(E_{y},-E_{x}).$ Here, one has adopted the following convection:
$\epsilon_{012}=\epsilon^{012}=\epsilon_{12}=\epsilon^{12}=1$, $F^{12}%
=F_{12}=-B$, $F_{0i}=E^{i}.$ As it is well-known, Bianchi identity corresponds
to the Faraday's law,%
\begin{equation}
\partial_{t}B+\nabla\times\mathbf{E}=0. \label{BI}%
\end{equation}
Eqs. (\ref{Elec1}),(\ref{Mag1}), (\ref{BI}) are the modified Maxwell equations
corresponding to Lagrangian (\ref{LEM}). Multiplying Eq. (\ref{Mag1}) by
$\epsilon_{ip},$ we have:%
\begin{equation}
(1+s)\partial_{p}B-\epsilon_{ip}\partial_{t}E^{i}+\epsilon_{ip}\left(
k_{DE}\right)  ^{ij}\partial_{t}E^{j}-\partial_{p}(L^{j}E^{j})-\epsilon
_{ip}L^{i}\partial_{t}B=\epsilon_{ip}J^{i}, \label{Mag2}%
\end{equation}
The stationary version of this equation is
\begin{equation}
n\partial_{i}B-2\partial_{i}(L^{j}E^{j})+L^{p}\partial_{p}E^{i}=-\epsilon
_{ip}J^{p}, \label{Mag3}%
\end{equation}
where $n=(1+s).$ Applying the operator $\partial_{i}$ on Eq.(\ref{Mag3}), it
turns out:
\begin{align}
n\nabla^{2}B-2\nabla^{2}(L^{j}E^{j})+L^{p}\partial_{p}\partial_{i}E^{i}  &
=-\epsilon_{ip}\partial_{i}J^{p},\label{Mag4}\\
n\nabla^{2}B+(L^{j}\partial_{j})\nabla^{2}A_{0}  &  =-\epsilon_{ip}%
\partial_{i}J^{p}. \label{Mag5}%
\end{align}
Multiplying Eq.(\ref{Elec1}) by $n$ and replacing Eq.(\ref{Mag3}) on it, it is
possible to achieve a decoupled expression for the electric field,%
\begin{equation}
\partial_{i}E^{i}-\left(  k_{DE}\right)  _{ij}\partial_{i}E^{j}+\frac{1}%
{n}L^{i}L^{q}\partial_{i}E^{q}=\rho+\frac{1}{n}\epsilon_{im}L^{i}J^{m}.
\label{Electric1}%
\end{equation}
The dependence on the current in the nonhomogeneous part indicates that this
planar model inherits a feature from the four-dimensional model: stationary
currents may engender both magnetic and electric fields. These modified
Maxwell equations exhibit an analogous form to the Maxwell ones of the
four-dimensional theory, respecting the structure of differential operators in
three and two spacial dimensions. A point of difference is that in the
stationary original theory, the coupling between the scalar and magnetic
sector is established only by the parity-odd coefficients. In this planar
theory, the coupling is implemented by the parity-odd and parity-even
coefficients $\left(  L^{i}\right)  $.

In the Lorentz gauge, $\partial_{\mu}A^{\mu}=0$, the wave equations for the
3-potential can be derived from
\begin{equation}
\square A^{\beta}-2Z^{\beta\alpha\lambda\kappa}\partial_{\alpha}%
\partial_{\lambda}A_{\kappa}=J^{\beta}. \label{Ami}%
\end{equation}
For $\beta=0$ and $\beta=i,$ we derive the equations for $A^{0}\ $and $A^{i}$,
namely:
\begin{equation}
\square A^{0}+\left(  k_{DE}\right)  ^{ij}\partial_{i}\partial_{j}A_{0}%
+L^{i}\partial_{i}B=\rho, \label{A0_1}%
\end{equation}%
\begin{equation}
\square A^{i}-\epsilon^{il}L^{j}\partial_{l}\partial_{0}A_{j}+\left(
k_{DE}\right)  ^{ij}\partial_{0}^{2}A_{j}-\left(  k_{DE}\right)  ^{ij}%
\partial_{0}\partial_{j}A_{0}+\epsilon_{il}\partial_{l}(L^{p}\partial
_{p})A_{0}+s\epsilon^{ip}\partial_{p}B-L^{i}\partial_{0}B=J^{i}, \label{A_1}%
\end{equation}
whose stationary versions are
\begin{equation}
\nabla^{2}A^{0}-\left(  k_{DE}\right)  ^{ij}\partial_{i}\partial_{j}%
A_{0}-L^{i}\partial_{i}B=-\rho, \label{A0_2}%
\end{equation}%
\begin{equation}
\nabla^{2}A^{i}-\epsilon_{il}\partial_{l}(L^{p}\partial_{p})A_{0}%
-s\epsilon^{ip}\partial_{p}B=-J^{i}. \label{A_2}%
\end{equation}
Using Eq.(\ref{Mag3}), the expression (\ref{A0_1}) for the scalar potential
can be decoupled as:
\begin{equation}
\lbrack\nabla^{2}-\left(  k_{DE}\right)  ^{ji}\partial_{i}\partial_{j}%
+\frac{1}{n}\left(  L^{j}L^{i}\partial_{i}\partial_{j}\right)  ]A_{0}%
=-\rho-\frac{1}{n}\epsilon_{im}L^{i}J^{m}. \label{A0_3}%
\end{equation}
This expression can be also obtained starting from the differential equation
for the electric field, Eq.(\ref{Electric1}), by replacing $E^{j}%
=-\partial_{j}A_{0}$. Considering that in the stationary regime it holds
$\partial_{i}A^{i}=0$, Eq.(\ref{A_2}) is written as
\begin{equation}
(1+s)\nabla^{2}A^{i}-\epsilon^{il}\partial_{l}(L^{p}\partial_{p})A_{0}=-J^{i}.
\end{equation}
This latter equation confirms that currents act as source for the both the
electric and magnetic fields. On the other hand, it is possible to show that
charges generate both electric and magnetic fields as well.

The magnetic field can be read from Eq. (\ref{Mag5}), $\nabla^{2}\left[
nB+(L^{j}\partial_{j})A_{0}\right]  =-\epsilon_{ip}\partial_{i}J^{p},$ leading
to
\begin{equation}
B\left(  \mathbf{r}\right)  =-\frac{1}{n}(L^{j}\partial_{j})A_{0}\left(
\mathbf{r}\right)  -\int G_{B}(\mathbf{r-r}^{\prime})[\frac{1}{n}\epsilon
_{im}\partial_{i}J^{m}(\mathbf{r}^{\prime})]d^{2}\mathbf{r}^{\prime},
\label{magnetic}%
\end{equation}
where the magnetic Green function satisfies
\begin{equation}
\nabla^{2}G_{B}(\mathbf{r-r}^{\prime})=\delta(\mathbf{r}-\mathbf{r}^{\prime}).
\end{equation}
In the momentum space, $\tilde{G}\left(  \mathbf{p}\right)  =-1/\mathbf{p}%
^{2},$ implying $G_{B}(\mathbf{r-r}^{\prime})=\frac{1}{2\pi}\ln\left\vert
\mathbf{r-r}^{\prime}\right\vert ,$ so that the magnetic field is written as
\begin{equation}
B\left(  \mathbf{r}\right)  =\frac{1}{n}L^{j}E^{j}\left(  \mathbf{r}\right)
-\frac{1}{2\pi}\frac{1}{n}\int\ln\left\vert \mathbf{r-r}^{\prime}\right\vert
[\epsilon_{im}\partial_{i}J^{m}(\mathbf{r}^{\prime})]d^{2}\mathbf{r}^{\prime}.
\label{B4}%
\end{equation}
In the absence of currents, a simple relation holds between the magnetic and
electric field:
\begin{equation}
B\left(  \mathbf{r}\right)  =\frac{1}{n}\left(  \mathbf{L\cdot E}%
(\mathbf{r})\right)  .
\end{equation}

An issue of interest is the complete wave equations which lead to the
dispersion relations of this planar electrodynamics. In order to study it, we
search for the wave equation for the electric field. We take the time derivate
of Eq. (\ref{Mag1}), and replace the Bianchi identity, $\partial_{t}B=-\left(
\epsilon_{mn}\partial_{m}E^{n}\right)  $ in it. \ After some manipulation, one
obtains a wave equation for the electric field at the form $M_{ij}E^{j}%
=0,$where
\begin{equation}
M_{ij}=[-n\partial_{j}\partial_{i}+n\delta_{ij}\nabla^{2}-[\delta_{ij}%
\partial_{t}^{2}-\left(  k_{DE}\right)  ^{ij}\partial_{t}^{2}]+L_{j}%
\partial_{t}\epsilon_{il}\partial_{l}-L_{i}\epsilon_{mj}\partial_{t}%
\partial_{m}].
\end{equation}
In the momentum space, it follows:
\begin{equation}
M_{ij}=[np_{j}p_{i}-n\delta_{ij}\mathbf{p}^{2}+[\delta_{ij}p_{0}^{2}-\left(
k_{DE}\right)  _{ij}p_{0}^{2}]+L_{j}\epsilon_{il}p_{0}p_{l}-L_{i}\epsilon
_{mj}p_{0}p_{m}].
\end{equation}
The dispersion relation is achieved imposing $\det\mathbb{M}=0.$ Evaluating
the components,
\begin{align}
M_{11} &  =[np_{1}^{2}-n\mathbf{p}^{2}+p_{0}^{2}-\left(  k_{DE}\right)
_{11}p_{0}^{2}+2L_{1}p_{0}p_{2}],\\
M_{22} &  =[np_{2}^{2}-n\mathbf{p}^{2}+p_{0}^{2}-\left(  k_{DE}\right)
_{22}p_{0}^{2}-2L_{2}p_{0}p_{1}],\\
M_{12} &  =M_{21}=np_{1}p_{2}-\left(  k_{DE}\right)  _{12}p_{0}^{2}+L_{2}%
p_{0}p_{2}-L_{1}p_{0}p_{1},
\end{align}
one can write and factor the determinant, $\det\mathbb{M}=M_{11}M_{22}-\left(
M_{12}\right)  ^{2},$ obtaining an exact dispersion relation:%
\begin{equation}
\det\mathbb{M=}p_{0}^{2}\left\{  \frac{{}}{{}}p_{0}^{2}\left[  1-\text{tr}%
(k_{DE})+\det(k_{DE})\right]  +2p_{0}[\mathbf{L\times p}+\left(
k_{DE}\right)  _{ij}p_{i}L_{j}^{\ast}]-[n\mathbf{p}^{2}-n\left(
k_{DE}\right)  _{ij}p_{i}p_{j}+(\mathbf{L\cdot p})^{2}]\right\}  =0.
\end{equation}
This physical dispersion relation yields the solution
\begin{equation}
p_{0}=\frac{1}{D}\left[  \mathbf{L\times p}+\left(  k_{DE}\right)  _{ij}%
p_{i}L_{j}^{\ast}\pm\Omega\left(  \mathbf{p}^{2}-\left(  k_{DE}\right)
_{ij}p_{i}p_{j}\right)  ^{1/2}\right]  .\label{DR2}%
\end{equation}
\textbf{where}%
\begin{align}
\mathbf{\ }D &  =[1-tr(k_{DE})+\det(k_{DE})],\\
\Omega &  =\left[  \left(  1+s\right)  D+\mathbf{L}^{2}-\left(  k_{DE}\right)
_{ij}L_{i}^{\ast}L_{j}^{\ast}\right]  ^{1/2}.
\end{align}
From relation (\ref{DR2}), we notice that both modes propagate with the same
phase velocity, which implies absence of birefringence. To understand it, we
should take the right $\left(  p_{0+}\right)  $ and the left $\left(
p_{0-}\right)  $ modes, corresponding to the $\pm$ signals in Eq.(\ref{DR2}),
propagating in the same sense. Hence, we should take the left $\left(
p_{0-}\right)  $ mode with reversed momentum\textbf{ (}$\mathbf{p\rightarrow
-p),}$
\begin{equation}
p_{0-}(-\mathbf{p)}=\frac{1}{D}\left[  -(\mathbf{L\times p)}-\left(
k_{DE}\right)  _{ij}p_{i}L_{j}^{\ast}-\Omega\left(  \mathbf{p}^{2}-\left(
k_{DE}\right)  _{ij}p_{i}p_{j}\right)  ^{1/2}\right]  .
\end{equation}
meaning propagation to the right. We should note that it coincides with the
right mode,
\begin{equation}
p_{0+}(\mathbf{p)}=\frac{1}{D}\left[  (\mathbf{L\times p)}+\left(
k_{DE}\right)  _{ij}p_{i}L_{j}^{\ast}+\Omega\left(  \mathbf{p}^{2}-\left(
k_{DE}\right)  _{ij}p_{i}p_{j}\right)  ^{1/2}\right]  ,
\end{equation}
with a reversed global signal. These relations provide the same phase
velocity. This situation is analogous to the one of the parity-odd dispersion
relation of the CPT-even original model, discussed in Eqs.(36-44) of
Ref.\cite{Prop}, which yields no birefringence.\ Such discussion reveals that
the six Lorentz-violating parameters of the electromagnetic sector, $s,$
$(k_{DE})_{ij}$, $L^{i}$, behave as nonbirefringent components (at any order).
Hence, the birefringent components of this planar theory should be contained
in the coupling tensor $T_{\mu\lambda\kappa}.$ As this tensor modifies the
equations of motion at second order, the birefringence is manifest only as a
second order effect in the Lorentz-violating parameters. At first order,
Eq.(\ref{DR2}) implies the following physical dispersion relation:%

\begin{equation}
p_{0}=\left\vert \mathbf{p}\right\vert [1+\frac{s}{2}+\frac{1}{2}%
\text{tr}(k_{DE})-\left(  k_{DE}\right)  _{ij}\frac{p_{i}p_{j}}{2\mathbf{p}%
^{2}}\pm\frac{\left(  \mathbf{L\times p}\right)  }{\left\vert \mathbf{p}%
\right\vert }]. \label{DR3}%
\end{equation}

From relation (\ref{DR3}), we can also evaluate the group velocity,%
\begin{equation}
u_{g}=[1+\frac{s}{2}+\frac{1}{2}\text{tr}(k_{DE})-\left(  k_{DE}\right)
_{ij}\frac{\hat{p}_{i}\hat{p}_{j}}{2}\pm\epsilon_{ij}L^{i}\hat{p}^{j}],
\end{equation}
showing that this electrodynamics could spoil causality. \ In order to perform
a complete analysis on the consistency of this theory (stability, causality,
and unitarity), one should carry out a detailed analysis via the Feynman gauge
propagator, which is being regarded as a future perspective.

\section{Classical Stationary Solutions}

In this section, we should solve the equations for the electromagnetic and
scalar sectors, obtaining stationary solutions at first order in the
Lorentz-violating parameters.

\subsection{The electrostatic and magnetostatic}

A good starting point to study the stationary solutions for the pure
electromagnetic sector is the differential equation for the scalar potential,
Eq. (\ref{A0_3}), which at first order is read as:
\begin{equation}
\lbrack\nabla^{2}-\left(  k_{DE}\right)  ^{ij}\partial_{i}\partial_{j}%
+\frac{1}{n}\left(  L^{j}L^{i}\partial_{i}\partial_{j}\right)  ]A_{0}%
=-\rho-\frac{1}{n}\epsilon_{im}L^{i}J^{m}.
\end{equation}
The solution for this equation can be achieved by means of the Green method,
which allows to write%
\begin{equation}
A_{0}\left(  \mathbf{r}\right)  =-\int G(\mathbf{r-r}^{\prime})[\rho
(\mathbf{r}^{\prime})+\frac{1}{n}\epsilon_{im}L^{i}J^{m}(\mathbf{r}^{\prime
})]d^{2}\mathbf{r}^{\prime}, \label{Green1A}%
\end{equation}
where $G(\mathbf{r}-\mathbf{r}^{\prime})$ is the\textbf{ }Green's function,
which fulfills the first order equation%
\begin{equation}
\lbrack\nabla^{2}-\left(  k_{DE}\right)  ^{ij}\partial_{i}\partial
_{j}]G(\mathbf{r}-\mathbf{r}^{\prime})=\delta(\mathbf{r}-\mathbf{r}^{\prime}).
\label{Green2}%
\end{equation}

In Fourier space it holds
\begin{equation}
G(\mathbf{r-r}^{\prime})=\frac{1}{\left(  2\pi\right)  ^{2}}\int
d^{2}\mathbf{p~}\tilde{G}\left(  \mathbf{p}\right)  \exp\left[  -i\mathbf{p}%
\cdot(\mathbf{r}-\mathbf{r}^{\prime})\right]  ,
\end{equation}
whose replacement in Eq.(\ref{Green2}) leads to
\begin{equation}
\tilde{G}\left(  \mathbf{p}\right)  =-\frac{1}{\mathbf{p}^{2}-\left(
k_{DE}\right)  ^{ji}p_{i}p_{j}}=-\frac{1}{\mathbf{p}^{2}}\left[  1+\left(
k_{DE}\right)  ^{ji}\frac{p_{i}p_{j}}{\mathbf{p}^{2}}+\ldots\right]  ,
\label{Green0}%
\end{equation}
and we have evaluated $\tilde{G}\left(  \mathbf{p}\right)  $ at first order in
the Lorentz-violating parameters, due to its usual smallness. Performing the
Fourier integrations, we achieve the\ following Green function:
\begin{equation}
G(\mathbf{R})=\frac{1}{\left(  2\pi\right)  }\left[  \left(  1+\frac{1}%
{2}\left(  k_{DE}\right)  ^{ii}\right)  \ln R+\frac{1}{2}\left(
k_{DE}\right)  ^{ij}\frac{R_{i}R_{j}}{R^{2}}\right]  \label{Green1}%
\end{equation}
where $\mathbf{R}=\left(  \mathbf{r}-\mathbf{r}^{\prime}\right)  \ $and the
terms involving the coefficients $\left(  k_{DE}\right)  ^{ij}$ are
corrections to usual planar Green function, $\ln R.$ Here, it were used the
following transforms:
\begin{equation}
\int d^{2}\mathbf{p~}\frac{1}{\mathbf{p}^{2}}e^{-i\mathbf{p}\cdot\mathbf{R}%
}=-2\pi\ln R,\text{ \ }\int d^{2}\mathbf{p~}\frac{p_{i}p_{j}}{\mathbf{p}^{4}%
}e^{-i\mathbf{p}\cdot(\mathbf{r}-\mathbf{r}^{\prime})}=-2\pi\left(
\frac{\delta_{ij}}{2}\ln R+\frac{1}{2}\frac{R_{j}R_{i}}{R^{2}}\right)  .
\end{equation}

The scalar potential is then written as%
\begin{equation}
A_{0}\left(  \mathbf{r}\right)  =-\frac{1}{2\pi}\int\left[  (1+\frac{1}%
{2}\left(  k_{DE}\right)  ^{ii})\ln\left\vert \mathbf{r-r}^{\prime}\right\vert
+\frac{1}{2}\left(  k_{DE}\right)  ^{ij}\frac{(\mathbf{r-r}^{\prime}%
)_{i}(\mathbf{r-r}^{\prime})_{j}}{(\mathbf{r-r}^{\prime})^{2}}\right]
[\rho(\mathbf{r}^{\prime})+\frac{1}{n}\epsilon_{im}L^{i}J^{m}(\mathbf{r}%
^{\prime})]d^{2}\mathbf{r}^{\prime},
\end{equation}
which at first order in the Lorentz-violating parameters is%
\begin{align}
A_{0}\left(  \mathbf{r}\right)   &  =-\frac{1}{2\pi}\int\left[  (1+\frac{1}%
{2}\left(  k_{DE}\right)  ^{ii})\ln\left\vert \mathbf{r-r}^{\prime}\right\vert
+\frac{1}{2}\left(  k_{DE}\right)  ^{ij}\frac{(\mathbf{r-r}^{\prime}%
)_{i}(\mathbf{r-r}^{\prime})_{j}}{(\mathbf{r-r}^{\prime})^{2}}\right]
\rho(\mathbf{r}^{\prime})d^{2}\mathbf{r}^{\prime}\\
&  -\frac{1}{2\pi}\epsilon_{im}L^{i}\int\ln\left\vert \mathbf{r-r}^{\prime
}\right\vert J^{m}(\mathbf{r}^{\prime})]d^{2}\mathbf{r}^{\prime},\nonumber
\end{align}
where $n^{-1}\sim\left(  1-s\right)  $. For a point-like charge distribution
($J^{m}(\mathbf{r}^{\prime})=0$), $\rho(\mathbf{r}^{\prime})=q\delta
(\mathbf{r}^{\prime}),$ one achieves the scalar potential
\begin{equation}
A_{0}\left(  \mathbf{r}\right)  =-\frac{q}{2\pi}\left[  [1+\frac{1}{2}\left(
k_{DE}\right)  ^{ii}]\ln r+\frac{1}{2}\left(  k_{DE}\right)  ^{ij}\frac
{r_{i}r_{j}}{r^{2}}\right]  . \label{A05}%
\end{equation}
This\ scalar potential differs from the usual planar behavior by the\ term
$\left(  k_{DE}\right)  ^{ij}r_{i}r_{j}/r^{2},$ which represents a directional
factor whose magnitude remains constant with distance. In fact, supposing
$r_{i}=r\cos\theta_{i},$ $r_{j}=r\cos\theta_{j},$ one has $\left(
k_{DE}\right)  ^{ij}r_{i}r_{j}/r^{2}=\left(  k_{DE}\right)  ^{ij}\cos
\theta_{i}\cos\theta_{j}.$ This shows that the Lorentz-violating corrections
are unable to modify the asymptotic behavior of the electric field.\ Hence,
the implied electric field,
\begin{align}
\mathbf{E}^{l}\left(  \mathbf{r}\right)   &  =\frac{q}{2\pi}\left[
[1+\frac{1}{2}\left(  k_{DE}\right)  ^{ii}]\frac{r^{l}}{r^{2}}+\frac{1}{r^{2}%
}\left(  \left(  k_{DE}\right)  ^{lj}r_{j}-\frac{\left(  k_{DE}\right)
^{ij}r_{i}r_{j}}{r^{2}}r^{l}\right)  \right]  ,\\
\mathbf{E}^{l}\left(  \mathbf{r}\right)   &  =\frac{q}{2\pi}\left[
[1+\frac{1}{2}\left(  k_{DE}\right)  ^{ii}-\left(  k_{DE}\right)  ^{ij}%
\cos\theta_{i}\cos\theta_{j}]\frac{r^{l}}{r^{2}}+\frac{1}{r^{2}}\left(
k_{DE}\right)  ^{lj}r_{j}\right]  ,
\end{align}
decays as $1/r$, as it occurs in the Maxwell theory in $(1+2)-$dimensions.
This field has a radial behavior except for the Lorentz-violating contribution
$\left(  k_{DE}\right)  ^{li}r_{i},$ which constitutes the qualitative
difference induced by Lorentz violation.

In this theory, a point-like charge, $[J^{m}(\mathbf{r}^{\prime})=0$,
$\rho(\mathbf{r}^{\prime})=q\delta(\mathbf{r}^{\prime})]$, yields a nonnull
magnetic field, that in accordance with Eq. (\ref{magnetic}), at first order
is $\ B\left(  \mathbf{r}\right)  =\left(  \mathbf{L\cdot E}(\mathbf{r}%
)\right)  .$ It then yields\textbf{ }%
\begin{equation}
B\left(  \mathbf{r}\right)  =\frac{q}{2\pi}\frac{\mathbf{L\cdot r}}{r^{2}}.
\end{equation}
This field decays with $1/r$ and does not present radial symmetry. It
possesses an angular dependence that reflects the direction of the vector
$\mathbf{r}$ in relation to the background vector $\mathbf{L.}$ In this case,
the modulation factor is $\left\vert \mathbf{L}\right\vert \cos\beta,$ being
$\beta$ the angle between $\mathbf{r}$ and $\mathbf{L.}$

A stationary current associated to a point-like charge with uniform velocity
$\mathbf{u}$, $\mathbf{J}(\mathbf{r}^{\prime})=q\mathbf{u}\delta
(\mathbf{r}^{\prime})$ , $\left[  \rho(\mathbf{r}^{\prime})=0\right]  $,
yields the scalar potential:
\begin{equation}
A_{0}\left(  \mathbf{r}\right)  =-\frac{q}{2\pi}\left(  \mathbf{L\times
u}\right)  \ln r,
\end{equation}
whose electric field is
\begin{equation}
E^{i}\left(  \mathbf{r}\right)  =\frac{q}{2\pi}\left[  \left(  \mathbf{L\times
u}\right)  \frac{r^{i}}{r^{2}}\right]  .
\end{equation}
Once the vector product engenders a scalar in two dimensions, this electric
field results aligned with the radial direction, without angular dependence.
The scalar $\left(  \mathbf{L\times u}\right)  $ acts as a modulation factor
sensitive to the angle between the vectors $\mathbf{L,u,}$ which can vary from
zero (for $\mathbf{L//u)}$ to the maximum $\left\vert \mathbf{L}\right\vert
\left\vert \mathbf{u}\right\vert $ (for $\mathbf{L\perp u).}$ The magnetic
field associated with this current density, $[\mathbf{J}(\mathbf{r}^{\prime
})=q\mathbf{u}\delta(\mathbf{r}^{\prime})$ , $\rho(\mathbf{r}^{\prime})=0]$,
is attained from Eq. (\ref{B4}),
\begin{equation}
B\left(  \mathbf{r}\right)  =-\frac{q}{2\pi}(1-s)\frac{\epsilon_{im}r_{i}%
u^{m}}{r^{2}}=\frac{q}{2\pi}(1-s)\frac{\mathbf{r\times u}}{r^{2}},
\end{equation}
where the Lorentz-violating contribution has the same form of Maxwell usual solution.

\subsection{The pure scalar sector}

A solution for the scalar field can be easily constructed. At first order, the
scalar field evolution is governed by Eq. (\ref{S2}), which in the stationary
limit is given by%
\begin{equation}
\left(  \nabla^{2}+2C^{ij}\partial_{i}{\partial}_{j}\right)  {\phi}{=}{J.}%
\end{equation}
he Green function for this equation satisfies $[\nabla^{2}+2C^{ij}\partial
_{i}\partial_{j}]G(\mathbf{r}-\mathbf{r}^{\prime})=\delta(\mathbf{r}%
-\mathbf{r}^{\prime}),$ while the solution is written as%
\begin{equation}
\phi\left(  \mathbf{r}\right)  =\int G(\mathbf{r-r}^{\prime})J(\mathbf{r}%
^{\prime})d^{2}\mathbf{r}^{\prime}.
\end{equation}
Following the procedure developed for the scalar field, we obtain
\begin{equation}
\tilde{G}\left(  \mathbf{p}\right)  =-\frac{1}{\mathbf{p}^{2}}\left[
1-2C^{ij}\frac{p_{i}p_{j}}{\mathbf{p}^{2}}\right]  ,
\end{equation}
the same structure Green's function for the scalar potential, Eq.
(\ref{Green0}). So, we attain as Green's function a result very similar to Eq.
(\ref{Green2}),
\begin{equation}
G(\mathbf{R})=\frac{1}{\left(  2\pi\right)  }\left[  \left(  1-C^{ii}\right)
\ln R-C^{ij}\frac{R_{j}R_{i}}{R^{2}}\right]  .
\end{equation}
The scalar field is given as
\begin{equation}
\phi\left(  \mathbf{r}\right)  =\frac{1}{2\pi}\int\left[  (1-C^{ii}%
)\ln\left\vert \mathbf{r-r}^{\prime}\right\vert -C^{ij}\frac{(\mathbf{r-r}%
^{\prime})_{i}(\mathbf{r-r}^{\prime})_{j}}{(\mathbf{r-r}^{\prime})^{2}%
}\right]  J(\mathbf{r}^{\prime})d^{2}\mathbf{r}^{\prime},
\end{equation}
The scalar field generated by a point-like scalar source, $J(\mathbf{r}%
^{\prime})=q\delta(\mathbf{r}^{\prime}),$ is
\begin{equation}
\phi\left(  \mathbf{r}\right)  =\frac{q}{2\pi}\left[  (1-C^{ii})\ln
r-C^{ij}\frac{r_{i}r_{j}}{r^{2}}\right]  .
\end{equation}
We thus confirm that scalar field presents a very similar behavior to the one
of the scalar potential, given by Eq. (\ref{A05}).

\section{Conclusions and remarks}

In this work, we have performed the dimensional reduction of the CPT-even
gauge sector of SME, attaining a planar model enclosing both gauge and scalar
sectors, coupled by a third-rank tensor stemming from the dimensional
reduction. The symmetries of the planar Lorentz-violating tensors have been
scrutinized, and the number of independent components were evaluated. The
parity of these components was determined considering the field $\phi$ as a
scalar, but it can be also examined supposing that $\phi$ behaves as a
pseudoscalar (inheriting the behavior of the component $A^{\left(  3\right)
})$. In the sequel, we have taken the coupling tensor as null, and examined
the equations of motion for the electromagnetic and scalar sectors. These
equations were solved by the Green's method in the stationary regime.

One parallel should be made with the dimensional reduction of the
Maxwell-Carroll-Field-Jackiw electrodynamics in Ref. \cite{Dreduction}. In
that case, it was obtained a planar model composed of the Maxwell-Chern-Simons
electrodynamics, a scalar field and a coupling term, where the Lorentz
violation was controlled by a 3-vector background, $v^{\mu}=($v$_{0}%
,$\textbf{v}$)$. The stationary classical solutions of this model revealed
that the background altered the asymptotic behavior of the fields. Indeed,
while the Maxwell-Chern-Simons solutions decay exponentially for
$r\rightarrow\infty,$ the Lorentz-violating solutions exhibited a $1/r$
\ behavior for $r\rightarrow\infty$. In the dimension reduction of the
CPT-even sector, the presence of Lorentz-violating terms do not alter the long
distance profile of the solutions, keeping the asymptotic behavior of the pure
Maxwell planar electrodynamics, $1/r$. It is noted, however, that the
Lorentz-violating parameters induce an angular dependence in the field
solutions. The canonical energy-momentum tensor was carried out, leading to an
energy density which is positive definite for small Lorentz-violating parameters.

The dispersion relation of the planar abelian gauge model was exactly
evaluated, revealing that the six Lorentz-violating parameters related to the
electromagnetic sector do not yield birefringence. This means that the pure
electrodynamics stemming from Lagrangian (\ref{Lag2}) presents no
birefringence at any order. Such effect, however, may be engendered by the
some components of the coupling tensor $T_{\mu\lambda\kappa},$ as a second
order effect. Finally, the group velocity evaluation shows that this planar
theory could be endowed with causality illness. A more careful analysis on the
physical consistency of this model (stability, causality, unitarity) is under
progress. Another point of interest is the investigation of topological
defects, such stable vortex configurations, in this theoretical framework.

\begin{acknowledgments}
The authors are grateful to FAPEMA, CAPES and CNPq (Brazilian research
agencies) for invaluable financial support. These authors also acknowledge J.
A. Helayel-Neto for interesting and useful discussions, and the IFT (Instituto
de F\'{\i}sica Te\'{o}rica) staff for the kind hospitality during the
realization of this work.
\end{acknowledgments}


\begin{thebibliography}{99}                                                                                               %


\bibitem {Colladay}D. Colladay and V. A. Kostelecky, \textit{Phys. Rev. }D
\textbf{55}, 6760 (1997); D. Colladay and V. A. Kostelecky, \textit{Phys. Rev.
}D \textbf{58}, 116002 (1998); S. R. Coleman and S. L. Glashow, \textit{Phys.
Rev}. D\textbf{\ 59}, 116008 (1999); S.R. Coleman and S.L. Glashow,
\textit{Phys. Rev}. D\textbf{\ 59}, 116008 (1999).

\bibitem {Samuel}V. A. Kostelecky and S. Samuel, \textit{Phys. Rev. Lett}.
\textbf{63}, 224 (1989); \textit{Phys. Rev. Lett}. \textbf{66}, 1811 (1991);
\textit{Phys. Rev. }D\textbf{\ 39}, 683 (1989); \textit{Phys. Rev.
}D\textbf{\ 40}, 1886 (1989), V. A. Kostelecky and R. Potting, \textit{Nucl.
Phys.} B\textbf{\ 359}, 545 (1991); Phys. Lett. B \textbf{381}, 89 (1996); V.
A. Kostelecky and R. Potting, \textit{Phys. Rev. }D\textbf{\ 51}, 3923 (1995).

\bibitem {Jackiw}S.M. Carroll, G.B. Field and R. Jackiw, Phys. Rev.\textit{\ }%
D \textbf{41}, 1231 (1990).

\bibitem {Higgs}A. P. Baeta Scarpelli, H. Belich, J. L. Boldo, J.A.
Helayel-Neto, Phys. Rev. D\textit{\ }\textbf{67}, 085021 (2003); A.P. Baeta
Scarpelli and J.A. Helayel-Neto, Phys.Rev. D \textbf{73}, 105020 (2006).

\bibitem {Adam}C. Adam and F. R. Klinkhamer, Nucl. Phys.\textit{\ }B
\textbf{607}, 247 (2001); C. Adam and F. R. Klinkhamer, Nucl. Phys.\textit{\ }%
B \textbf{657}, 214 (2003).

\bibitem {Soldati}A.A. Andrianov and R. Soldati, Phys. Rev. D \textbf{51},
5961 (1995); Phys. Lett. B \textbf{435}, 449 (1998); A.A. Andrianov, R.
Soldati and L. Sorbo, Phys. Rev. D \textbf{59}, 025002 (1998); A. A.
Andrianov, D. Espriu, P. Giacconi, R. Soldati, J. High Energy Phys. 0909, 057
(2009); J. Alfaro, A.A. Andrianov, M. Cambiaso, P. Giacconi, R. Soldati,
Int.J.Mod.Phys.A 25, 3271 (2010); V. Ch. Zhukovsky, A. E. Lobanov, E. M.
Murchikova, Phys.Rev. D73 065016, (2006).

\bibitem {Susy}M.S. Berger, V. A. Kostelecky, Phys.Rev.D 65, 091701 (2002); H.
Belich , J.L. Boldo, L.P. Colatto, J.A. Helayel-Neto, A.L.M.A. Nogueira,
Phys.Rev. D \textbf{68}, 065030 (2003); A.P. Baeta Scarpelli, H. Belich, J.L.
Boldo, L.P. Colatto, J.A. Helayel-Neto, A.L.M.A. Nogueira, Nucl. Phys. Proc.
Suppl.127, 105-109 (2004).

\bibitem {Radio}R. Jackiw and V. A. Kosteleck\'{y}, Phys. Rev. Lett.
\textbf{82}, 3572 (1999); J. M. Chung and B. K. Chung Phys. Rev.
D\textbf{\ 63}, 105015 (2001); J.M. Chung, Phys.Rev. D\ \textbf{60}, 127901
(1999); G. Bonneau, Nucl.Phys. B\ \textbf{593}, 398 (2001); M. Perez-Victoria,
Phys. Rev. Lett. \textbf{83}, 2518 (1999); M. Perez-Victoria, J. High. Energy
Phys. \textbf{\ 0104}, (2001) 032; O.A. Battistel and G. Dallabona, Nucl.
Phys. B \textbf{610}, 316 (2001); O.A. Battistel and G. Dallabona, J. Phys. G
\textbf{28}, L23 (2002); J. Phys. G \textbf{27}, L53 (2001); A. P. B.
Scarpelli, M. Sampaio, M. C. Nemes, and B. Hiller, Phys. Rev. D\ \textbf{64},
046013 (2001); T. Mariz, J.R. Nascimento, E. Passos, R.F. Ribeiro and F.A.
Brito, J. High. Energy Phys. \textbf{0510} (2005) 019; F.A. Brito, J.R.
Nascimento, E. Passos, A. Petrov, J. High. Energy Phys. \textbf{0706} (2007)
016; B. Altschul, Phys. Rev. D \textbf{70}, 101701 (2004); A.P.B. Scarpelli,
M. Sampaio, M.C. Nemes, B. Hiller, Eur. Phys. J. C \textbf{56}, 571 (2008);
Oswaldo M. Del Cima, J. M. Fonseca, D.H.T. Franco, O. Piguet, Phys. Lett.
\textbf{B} 688, 258 (2010); T. Mariz, Phys.Rev. D \textbf{83}, 045018 (2011).

\bibitem {Cerenkov1}R. Lehnert and R. Potting, Phys. Rev. Lett. \textbf{93},
110402 (2004); R. Lehnert and R. Potting, Phys. Rev. D \textbf{70}, 125010
(2004); B. Altschul, Phys. Rev. D \textbf{75}, 105003 (2007); C. Kaufhold and
F.R. Klinkhamer, Nucl. Phys. B \textbf{734, 1 }(2006).

\bibitem {winder2}A.H. Gomes, J.M. Fonseca, W.A. Moura-Melo, A.R. Pereira,
JHEP 05, 104 (2010).

\bibitem {Casimir}M. Frank and I. Turan, Phys. Rev. \textbf{D 74}, 033016
(2006); O.G. Kharlanov, V.Ch. Zhukovsky, Phys. Rev. \textbf{D 8}1, 025015 (2010).

\bibitem {winder1}J. M. Fonseca, A. H. Gomes, W. A. Moura-Melo, Phys. Lett. B
\textbf{671}, 280 (2009).

\bibitem {Tfinite}R. Casana, M. M. Ferreira Jr. and J. S. Rodrigues, Phys.
Rev. D \textbf{78}, 125013 (2008).

\bibitem {CMBR}J.-Q. Xia, Hong Li, X. Wang, X. Zhang, Astron.Astrophys. 483,
715 (2008); J.-Q. Xia, H. Li, X. Zhang, Phys.Lett. B \textbf{687}, 129 (2010);
B. Feng, M. Li, J.-Q. Xia, X. Chen, X. Zhang, Phys. Rev. Lett. \textbf{96},
221302 (2006); P. Cabella, P. Natoli, J. Silk, Phys. Rev. D \textbf{76},
123014 (2007).

\bibitem {Classical}R. Casana, M.M. Ferreira, and C.E.H. Santos, Phys. Rev. D
\textbf{78}, 105014 (2008).

\bibitem {Dreduction}H. Belich, M.M. Ferreira Jr., J.A. Helayel-Neto, M.T.D.
Orlando, Phys. Rev. D \textbf{67},125011 (2003); Erratum-ibid., Phys. Rev.
\textbf{D 69}, 109903 (2004).

\bibitem {Manojr2}H. Belich, M.M. Ferreira Jr., J.A. Helayel-Neto, M.T.D.
Orlando, Phys. Rev. D \textbf{68}, 025005 (2003).

\bibitem {Dreduction2}H. Belich, M.M. Ferreira Jr., J.A. Helayel-Neto, Eur.
Phys. J. C 38, 511 (2005).

\bibitem {KM1}V. A. Kostelecky and M. Mewes, Phys. Rev. Lett. \textbf{87},
251304 (2001).

\bibitem {KM2}V. A. Kostelecky and M. Mewes, Phys. Rev. D\textbf{\ 66}, 056005 (2002).

\bibitem {KM3}V. A. Kostelecky and M. Mewes, Phys. Rev. Lett. \textbf{97},
140401 (2006).

\bibitem {Cherenkov2}B. Altschul, Nucl. Phys. B\textbf{\ 796}, 262 (2008); B.
Altschul, Phys. Rev. Lett. \textbf{98}, 041603 (2007); C. Kaufhold and F.R.
Klinkhamer, Phys. Rev. D \textbf{76}, 025024 (2007)\textbf{. }

\bibitem {Klink2}F.R. Klinkhamer and M. Risse, Phys. Rev. D \textbf{77},
016002 (2008); F.R. Klinkhamer and M. Risse, Phys. Rev. D \textbf{77}, 117901
(A) (2008).

\bibitem {Klink3}F. R. Klinkhamer and M. Schreck, Phys. Rev. D \textbf{78},
085026 (2008).

\bibitem {Interac1}V. A. Kostelecky and A.G.M. Pickering, Phys. Rev. Lett.
\textbf{91}, 031801 (2003); B. Altschul, Phys.Rev. D \textbf{70, }056005 (2004).

\bibitem {Interac2}C.D. Carone, M. Sher, and M. \ Vanderhaeghen, Phys. Rev. D
\textbf{74}, 077901 (2006); B. Altschul, Phys. Rev. D \textbf{79}, 016004 (2009).

\bibitem {Interac3}M.A. Hohensee, R. Lehnert, D. F. Phillips, R. L. Walsworth,
Phys. Rev. D \textbf{80}, 036010(2009); M.A. Hohensee, R. Lehnert, D. F.
Phillips, R. L. Walsworth, Phys. Rev. Lett. \textbf{102}, 170402 (2009); B.
Altschul, Phys. Rev. D \textbf{80}, 091901(R) (2009).

\bibitem {Bocquet}J.-P. Bocquet \textit{et at.}, Phys. Rev. Lett.
\textbf{104,} 241601 (2010).

\bibitem {LV}M.N. Barreto, D. Bazeia, R. Menezes, Phys.Rev.\textbf{ D 73},
065015 (2006); D. Bazeia, M. M. Ferreira Jr., A. R. Gomes, R. Menezes, Physica
\textbf{D 239}, 947 (2010); A. de Souza Dutra, R. A. C. Correa, Phys. Rev. D
\textbf{83}, 105007 (2011).

\bibitem {Brito}M. A. Anacleto, F. A. Brito, E. Passos, Phys. Lett. B\textbf{
694}, (2010); M. A. Anacleto, F. A. Brito, E. Passos, arXiv:1101.2891.

\bibitem {Casana}R Casana, K. A. T. da Silva, arXiv:1106.5534.

\bibitem {Planar}S. Deser, R. Jackiw, and S. Templeton, Ann. Phys. (NY) 140,
372 (1982); Gerald V. Dunne, arXiv:hep-th/9902115.

\bibitem {Prop}R. Casana, M. M. Ferreira Jr, A. R. Gomes, P. R. D. Pinheiro,
Phys. Rev. D \textbf{80}, 125040 (2009).
\end{thebibliography}
\end{document}